%% file: main.tex
\documentclass[sigconf,authorversion,nonacm]{acmart}

\AtBeginDocument{%
  \providecommand\BibTeX{{%
    \normalfont B\kern-0.5em{\scshape i\kern-0.25em b}\kern-0.8em\TeX}}}

\usepackage{txfontsb}

\usepackage{graphicx}
\usepackage{soul}
\usepackage{listings}
\usepackage{color}
\usepackage{adjustbox}
\usepackage{balance}

\usepackage{pifont}
\usepackage{amssymb}
\usepackage{rotating}
\usepackage{multirow}
\usepackage{tikz}
\usepackage{amsmath}
\usepackage{xspace}
\usepackage{multirow}
\usepackage{booktabs}

\usepackage{array}
\newcolumntype{L}[1]{
  >{\raggedright\let\newline\\\arraybackslash\hspace{0pt}}m{#1}}
\newcolumntype{C}[1]{
  >{\centering\let\newline\\\arraybackslash\hspace{0pt}}m{#1}}
\newcolumntype{R}[1]{
  >{\raggedleft\let\newline\\\arraybackslash\hspace{0pt}}m{#1}}

\usepackage{graphicx,subcaption}

\usepackage{makecell}
\renewcommand\footnotetextcopyrightpermission[1]{}

\newcommand{\system}{IDWork\xspace}

\newcommand{\iot}{IoT\xspace}

\AtBeginDocument{%
  \providecommand\BibTeX{{%
    \normalfont B\kern-0.5em{\scshape i\kern-0.25em b}\kern-0.8em\TeX}}}


\begin{document}

\title{Position paper: A systematic framework for categorising \iot device fingerprinting mechanisms}

\author{Poonam Yadav}
\affiliation{University of York, UK}

\author{Angelo Feraudo}
\affiliation{University of York, UK}
\author{Budi Arief}
\affiliation{University of Kent, UK}

\author{Siamak F. Shahandashti}
\affiliation{University of York, UK}
\author{ Vassilios G. Vassilakis}
\affiliation{University of York, UK}

\renewcommand{\shortauthors}{P. Yadav et al.}


\begin{abstract}
The popularity of the Internet of Things (IoT) devices makes it increasingly important to be able to fingerprint them, for example in order to detect if there are misbehaving or even malicious IoT devices in one's network. However, there are many challenges faced in the task of fingerprinting IoT devices, mainly due to the huge variety of the devices involved. At the same time, the task can potentially be improved by applying machine learning techniques for better accuracy and efficiency.
The aim of this paper is to provide a systematic categorisation of machine learning augmented techniques that can be used for fingerprinting IoT devices. This can serve as a baseline for comparing various IoT fingerprinting mechanisms, so that network administrators can choose one or more mechanisms that are appropriate for monitoring and maintaining their network.
We carried out an extensive literature review of existing papers on fingerprinting IoT devices -- paying close attention to those with machine learning features. This is followed by an extraction of important and comparable features among the mechanisms outlined in those papers.
As a result, we came up with a key set of terminologies that are relevant both in the fingerprinting context and in the IoT domain. This enabled us to construct a framework called \textbf{\textit{\system}}, which can be used for categorising existing IoT fingerprinting mechanisms in a way that will facilitate a coherent and fair comparison of these mechanisms.
We found that the majority of the IoT fingerprinting mechanisms take a passive approach -- mainly through network sniffing -- instead of being intrusive and interactive with the device of interest. Additionally, a significant number of the surveyed mechanisms employ both static and dynamic approaches, in order to benefit from complementary features that can be more robust against certain attacks such as spoofing and replay attacks.
\end{abstract}


\maketitle


\input{1_introduction.tex}
\input{2_background+related.tex}
\input{3_methods.tex}
\input{4_fingerprinting.tex}
\input{5_conclusion.tex}


\bibliographystyle{ACM-Reference-Format}
\balance
\bibliography{references}

\end{document}

%% file: 1_introduction.tex
\section{Introduction}
\label{sec:introduction}
Device fingerprinting is a process of identifying a device or device type based on its unique intrinsic or behavioural properties~\cite{xu2015device, gu2018fingerprinting-hindawi, hamad2019iot}. Device fingerprinting is very popular in internet-connected general purpose computing devices to track user behaviour and application usage. Some of the interesting applications include browser fingerprinting for web analytics, user tracking, fraud detection and accountability~\cite{Laperdrix2020, Hupperich2015, Crabtree2018} and have gained a significant interests from cyber-security community. While it is clear that device fingerprinting can bring a lot of benefits -- especially for providing automated and customisable user experience -- there are also concerns that it can pose security and privacy risks \cite{manginointernet, ren2019information}. On the other hand, it has been suggested that device fingerprinting can also be used to help improve the security of smart home automation \cite{jose2016improving}.

There are three main properties that need to be satisfied in order to achieve effective fingerprinting of devices \cite{xu2015device}:
\begin{itemize}
\item \textit{unique identity property}: the device fingerprint has to be unique for the device;
\item \textit{integrity property}: the fingerprints must be impossible -- or at least, difficult -- to forge;
\item \textit{reproducibility property}: the features used in the fingerprinting process must be stable, especially in the presence of environment changes and mobility.
\end{itemize}



The increased prevalence of Internet of Things (IoT) devices makes the task of fingerprinting devices more challenging. To start with, there is a great variety of IoT devices available on the market, and there are many different ways for them to operate. These pose a challenge in creating a generic mechanism that can perform accurate fingerprinting of all IoT devices. Furthermore, there are some fundamental differences between IoT devices and general computers (for which, more mature fingerprinting mechanisms have been developed). For example,  IoT devices do not have many standard browser-based applications, which means many standard fingerprinting mechanisms will not work for IoT devices. Moreover, many IoT devices do not have a standard Graphical User Interface~(GUI) and they might even work autonomously in pervasive environments without user's direct control. 

Due to resource constraints and insecure designs, IoT devices are prone to be involved in cyber-attacks, ranging from being the target \cite{skowron2020traffic, Hariri2019} to being exploited to create a botnet to mount a massive Distributed Denial of Service attack \cite{Antonakakis2017, mirai}. Therefore, it is necessary and desirable to be able to automatically detect whether certain IoT devices might be vulnerable or could be exploited in cyber-attacks. The automatic device identification is one of the core requirements for building a secure IoT ecosystem, including cyber-attack and anomaly detection systems and automatic management and control. 

Various device fingerprinting mechanisms have been proposed in the last few years. However, not all of these mechanisms are suitable for the IoT domain. Many IoT device fingerprinting mechanisms are only suitable for specific use cases or tailored to certain requirements, making it challenging to choose a correct fingerprinting mechanism that will be appropriate for a new use case, for example. This shortcoming is the motivation behind our paper. In this paper, we explore and collate existing IoT fingerprinting mechanisms -- especially those that leverage Machine Learning (ML) techniques -- and present a holistic view and terminologies used in the fingerprinting context, which can be used for further research and development.

\vspace{0.2cm}
\noindent
{\bf{Contributions.}} The key contributions of our paper are:
\begin{itemize}
    \item The compilation of a key set of fingerprinting terminologies.
    \item The identification of important features to be included for achieving effective and accurate fingerprinting of IoT devices.
    \item The construction of \textbf{\textit{\system}}: a systematic framework for categorising IoT device fingerprinting mechanisms, which can be used for comparing and selecting suitable fingerprinting mechanism(s) for an IoT application.
\end{itemize}

The rest of the paper is organised as follows. Section~\ref{sec:background} provides some background and related work, while Section~\ref{sec:methods} outlines the methodology we followed in our research. Section \ref{sec:fingerprinting} represents the core of our work, giving an overview of the \textbf{\textit{\system}} framework, along with the key terminologies and our results. Finally, Section~\ref{sec:conclusion} concludes our paper and provides some ideas for future work.

%% file: 2_background+related.tex
\section{Related Work}
\label{sec:background}

Several papers have discussed various fingerprinting mechanisms, although they are not necessarily dedicated to IoT devices~\cite{xu2015device, Franklin2006, Wang2018}. Cunche et al. \cite{cunche2014linking} looked into device fingerprinting based on monitoring wireless probes that a device may make, based on the preferred network (access points) stored on that device. The main concern of the paper was privacy infringement, for example by exploiting information contained in the fingerprint to infer social links between device owners.
Spooren et al. \cite{spooren2015mobile} provides a critical assessment of device fingerprinting for risk-based authentication. In particular, they pointed out that device fingerprinting carries a lot of similarity among mobile devices, making this approach less reliable for risk assessment and step-up authentication.

Ferrag et al. \cite{ferrag2019authentication} looked at human physiological and behavioural features in their investigation into factors that might hinder biometrics models’ development and deployment on a large scale for authenticating IoT devices. They classified related survey papers based on deployment scope, focus biometric area, threat models, countermeasures, as well as the ML algorithms and Data Mining methods used by existing authentication and authorisation schemes for IoT devices. The paper also listed a set of biofeatures that can be used for biometric authentication of IoT devices, including gaze gestures, electrocardiogram, keystroke dynamics, fingerprint, ear shape and hand geometry \cite{ferrag2019authentication}. They focus only on biofeatures, so other traits (such as network characteristics and device information) are not considered.


Skowron et al. \cite{skowron2020traffic} study the information leakage exposed by traffic fingerprinting attacks. They use features of statistical network flows and ML. Hence, this approach is effective even when the IoT traffic is encrypted. This approach relies on decision trees (CART classifier) and heuristically creates random forests out of 100 trees. The proposed approach aims at both device identification as well as the detection of anomalous user activities. The latter is based on features such as packet size, packet inter-arrival time, and transmission rate. 

Hamad et al. \cite{hamad2019iot} perform IoT device identification using traffic characteristics (network flows), based on real-time devices connected to an IP network. A passive behavioural fingerprinting approach is used, while device classification is based on features extracted from both packet header and payload. These include IP address, packet size, and other traffic related features. The authors investigated different supervised learning classifiers such as ABOOST, LDA, KNN, Decision Tree, Naive Bayesian, and SVM Random Forest (RF), with RF showing the best performance.






%% file: 3_methods.tex
\section{Methods}
\label{sec:methods}



Between March and June 2020, we carried out a systematic review of relevant papers that have been published on various venues, including USENIX, ACM, IEEE, Nature, ScienceDirect, MDPI, Springer, Elsevier, and Hindawi. We also utilised Google Scholar for comparison and for augmenting our search. The following keywords were used to gather the initial set of papers: ``IoT'', ``fingerprinting'', ``device identification'', ``device authentication'', ``device authorisation'', ``traffic filtering'', ``anomaly detection''. 

We then divided up these papers among ourselves in order to analyse them and to extract key features of each paper. In order to be able to compare these papers fairly and consistently, we constructed a framework we call {\bf {\em {\system}}}, as outlined in Section \ref{sec:framework}. At various stages, we also performed synchronisation checks among ourselves to make sure the process is robust and consistent. 

Further papers were added to the list between July and September 2020, mostly as a result of following up cited papers by those in the initial main list. The same analysis and extraction process using {\bf {\em {\system}}} were applied to these additional papers to ensure consistency.

Our process came up with a final list of 31 articles from the 142 papers that we analysed carefully. The papers in our final list are shown in Table \ref{tab:classification}, which is constructed by populating a table as a result of applying our {\bf {\em {\system}}} framework. 

%% file: 4_fingerprinting.tex
\section{Overview of \system}\label{sec:fingerprinting}
\begin{figure}
    \includegraphics[width=\columnwidth]{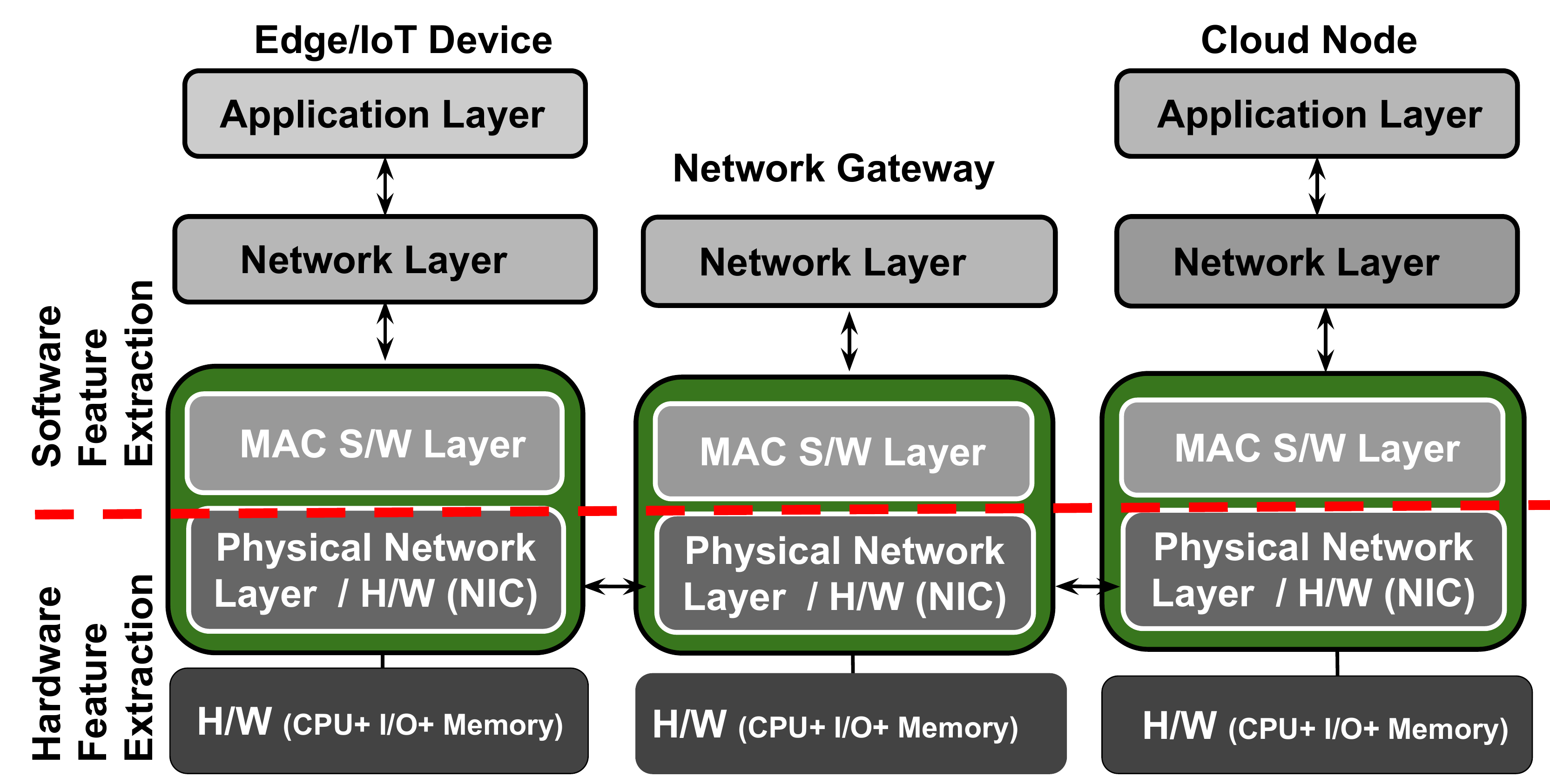}
    \caption{Edge/IoT device network end-to-end components}
    \label{fig:iot_block}
\end{figure}

For building {\bf {\em {\system}}}, we systematically reviewed the literature and recent state-of-the-art work to understand different terminologies used in IoT fingerprinting mechanisms. We simplified the presentation of IoT end-to-end ecosystem, as shown in Figure~\ref{fig:iot_block}. The Edge/IoT device -- showing the TCP/IP networking stack -- is partitioned horizontally. The upper partition represents \textit{software feature extraction}, which is composed of  application, networking and Medium Access (MAC) layers. The lower partition represents \textit{hardware feature extraction} comprising of two sub-layers - the first layer (upper) represents the features extracted from the Physical Network Layer along with the firmware and hardware (Networking Interface responsible for Link/Physical layer communication) layers and the second layer (lower) represents the hardware features which explicitly use the  hardware device.  Similarly, same  layer partition is performed at Network Gateway and Cloud Node. 


\subsection{Fingerprinting terminology}\label{sec:term}
In this section, we explain the key terminology that is important to grasp, before we define the \textbf{\textit{\system}} framework.\\\\
\textbf{Passive vs Active Fingerprinting}: In passive fingerprinting, we collect information produced by a device and create an identification pattern by only observing the data coming from the device, i.e., no interaction with the device is carried out. In active fingerprinting, we instigate the target device to produce useful information, e.g. by making the device emit particular signals (at the physical-layer), or by producing packets which require a specific response from the device (at the network-layer). Thus, the difference between the two methods is that the former uses a sniffer to capture and analyse traffic, but it does not send traffic to the target~\cite{yang2019towards}, while the latter sends queries to the target and analyses the response.\\\\
\textbf{Static Feature vs Dynamic Feature Fingerprinting}: A static feature fingerprinting includes only features that do not usually change over the time (e.g. MAC address), while dynamic feature fingerprinting uses dynamic features that can change over time such as inter-arrival time associated with data flow. \\\\
\textbf{Adaptive vs Fixed Fingerprinting Algorithms}: An ``adaptive'' fingerprinting approach uses an algorithm that changes in response to certain conditions observed during the fingerprinting process. On the other hand, when the fingerprinting process always uses the same (deterministic) algorithm with pre-determined and constant parameters for all cases observed, we can consider that as ``fixed''.\\\\
\textbf{Hardware Feature vs Software Feature Fingerprinting}: The former approach uses features that are extracted using special Physical Unclonable Function (PUF) circuits to capture hardware-intrinsic properties. The latter approach uses behavioural software properties, which could be found in the network traffic generated by the IoT device.\\\\
\textbf{Rule-based vs ML-based Fingerprinting}: In rule-based approaches, the fingerprinting criteria are mathematically-formalised in the form of {\em {if-then-else}} rules. Such rules are often defined by thresholds and used to create fingerprints by capturing the correlation between the features/parameters. In ML-based fingerprinting, an ML model is created using the input features/parameters, and trained on the data to learn possible data correlations for generalisation.\\\\
\textbf{Device vs Network vs Cloud Level}: Fingerprinting approaches can act on different levels. Usually in the case of device-level fingerprinting, the approach generates a device signature which relies on its hardware characteristics, e.g. Radio Signal or clock skew. When the approach analyses network traffic -- i.e. there is an additional entity within the network that monitor the traffic to produce device pattern -- we refer to it as network-level fingerprinting. It is even possible that fingerprinting procedures are applied externally to a network so that they can be executed on multiple networks. We refer to this case as cloud-level figerprinting.\\\\
\textbf{White-box vs Black-box Fingerprinting}: White-box fingerprinting is possible when we can directly access a device's firmware source code and then build a dynamic model of that device~\cite{Li2018}. Black-box fingerprinting exploits the interaction between different layers (e.g. application layer and transport layer) to build devices' fingerprints.\\\\
\textbf{Unique Device Identification vs Type Identification vs Class Identification}: Fingerprinting approaches can have different outputs depending on the designer's goal. In particular they can produce: a unique device identifier, device model or device class (devices with similar properties). \\\\
\textbf{Supervised vs Unsupervised ML-based Fingerprinting}: Supervised learning involves labeled data, which means that a prior knowledge about the classification of the learning data is provided. Conversely, unsupervised learning involves unlabeled data, so the ML goal is to infer a suitable classification of the data involved as well as classifying the data.\\\\
\textbf{Radio vs MAC vs Network vs Application Layer}:  Radio fingerprinting exploits the unique characteristics in the radio signal emitted by a device. MAC fingerprinting exploits the characteristics of the data frames produced by a device (e.g. probe request in Wi-Fi). In network fingerprinting, the network packet parameters are used to build an identification pattern. Application fingerprinting approaches typically gather information to find out the device's services and operating system. \\\\
\textbf{Open-world vs Close-world Evaluation}: Open-world refers to any approach that is able to identify IoT devices within a larger set of devices not only restricted to IoT devices. Closed-world is when identification is evaluated on data that is restricted to only IoT devices. \\\\
\textbf{Network Packet vs Flow-based Features}: A fingerprinting approach that relies on network traffic can use packet-based or flow-based features, or both. Packet-based features use the content of individual packet payloads and headers, whereas flow-based features are based on temporal features of multiple packets coming from the same device, e.g. packet flow direction, inter-arrival time and inter-packet length~\cite{Moustafa2015}.\\\\
\textbf{Network Packet Header vs Deep Packet Based Features}: When fingerprinting involves packet payload we refer to it as using deep-packet features. Otherwise, when only packet header parameters are used to build an identification pattern, we refer to it as using packet header features.\\\\
\textbf{Encrypted vs Unencrypted Network Traffic}: 
Some fingerprinting approaches do not need access to the packet payload, i.e. they can work on both encrypted and unencrypted packets. Conversely, others are designed to work on encrypted and unencrypted packet payloads, such as the algorithm proposed by Robyns et al.~\cite{robyns2017noncooperative} which exploits per-bit entropy analysis (MAC address randomization). Furthermore, some approaches are able to extract the features required only if the payload is not encrypted. \\\\
\textbf{Real Devices vs Simulated}: Fingerprinting approaches based on deep-learning algorithms require a large amount of data to properly identify devices. Moreover, budget constraints do not allow for a large number of devices to produce an exhaustive dataset for evaluation purposes. Thus, tools to simulate IoT devices are used (e.g. Node-RED \cite{koroniotis2019towards}). In this case, traffic flows and most of the important features of typical categories of IoT devices -- such as fridges or washing machines -- can be simulated and used to build  datasets. On the other hand, if real devices are used, the results will be more representative of real-world scenarios, but the budgetary requirements are higher as well.\\\\
\textbf{Testbed vs Real World Evaluation}: Fingerprinting approaches are developed either on testbeds or in real-world environments. The latter approaches provide additional resilience and deployment credibility for the fingerprinting algorithm. 


\input{mtable}

\subsection{\system Framework}\label{sec:framework}
\begin{figure}[ht]
    \includegraphics[width=1\columnwidth]{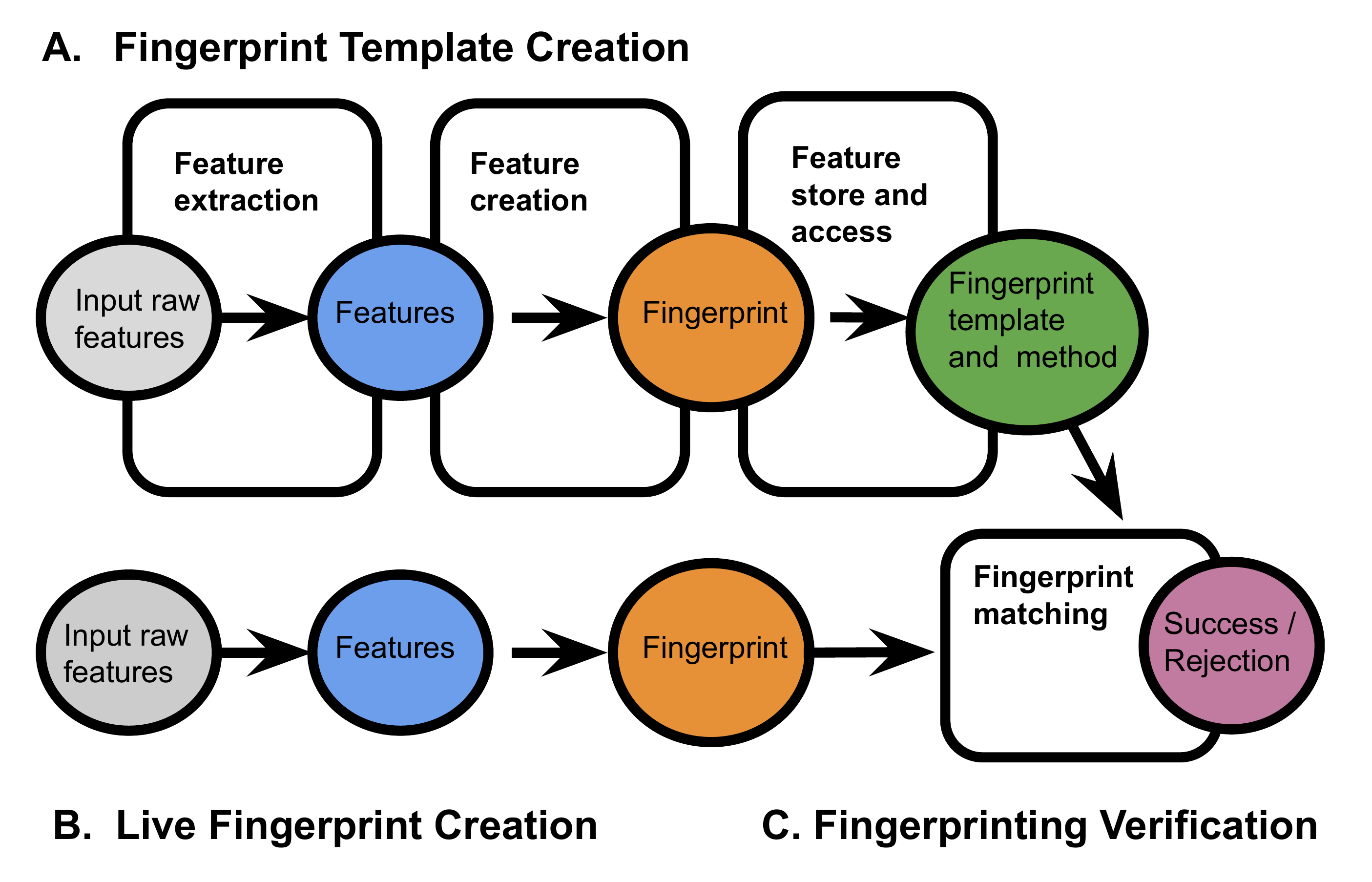}
    \caption{IoT device fingerprinting workflow.}
    \label{fig:iot_fingerprint}
\end{figure}

The construction of {\bf {\em {\system}}} started with an understanding of the basic fingerprinting creation and verification workflow, as shown in Figure~\ref{fig:iot_fingerprint}. The fingerprinting process consists of three steps: (1) Fingerprint template creation and storage, (2) Live fingerprint creation, and (3) Fingerprint verification. We analysed different considerations under each step. The task of initial template creation is a one-time process. The live fingerprint creation process may or may not follow the same approach or steps; however, the general approach follows these two sub-steps every time: feature extraction from the raw input features, and fingerprint generation.
In the fingerprint template creation, the final step involving template storage and access mechanism is essential.  In our work, we have considered different options under each sub-step (as shown in Figure~\ref{fig:iot_fingerprint}), that differentiates one fingerprinting mechanism from another.  Some of these options are either implementation-related or real-time deployment-related and they need to be considered, analysed and accessed before deciding an appropriate fingerprinting process suitable for an IoT application. Some of the important options we explored are:
\begin{itemize}
\item [--] Does fingerprint feature extraction, or creation mechanism need device access to gather fingerprinting raw input?
\item [--] Does fingerprint feature extraction, or creation mechanism need invasive mechanisms?
\item [--] Does fingerprint feature extraction, or creation mechanism need additional hardware?
\item [--] What is the fingerprint feature extraction and creation, or what is the storage location in the IoT end-to-end system, for example,  on the device, at a network gateway or on a cloud server? 
\item [--] What are the security, integrity, anti-tempering considerations when storing and accessing the pre-created fingerprints?
\item [--] What is the computational complexity of individual steps?
\end{itemize}

After careful consideration, we picked seven labels categorised under three broad steps: \textit{fingerprinting methods}, \textit{fingerprinting input features} and \textit{fingerprinting output}.  
We studied two categorisations under \textit{fingerprinting methods}, namely Passive (P)/Active (A) fingerprinting and Static (S)/Dynamic (D) fingerprinting. 
Under \textit{fingerprinting input features}, we considered categorisation using TCP/IP networking stack label and we used three choices: MAC/Network/Application Layers under this label. We further analysed these layers with two sub-labels with these choices: Packet/Dataframe/Flow and Header/Payload.  We also explored the mechanisms which use these sub-labels for fingerprint creation, and we categorise them as Rule-based or ML-based.
We classify \textit{fingerprint output} into three categories: Class, Type and Unique.

As a summary, in this work we present seven important features (as shown in Table~\ref{tab:classification}), which broadly allow for a systematic and consistent way for classifying IoT fingerprinting mechanisms into logical categories. In total, there are potentially 432 exhaustive combinations, but certain combinations are more prevalent than others. We perceive these popular categories as the more promising approaches that one should take in their effort to achieve a meaningful fingerprinting exercise of IoT devices.

\footnotetext[1]{They use an ML algorithm only for evaluation purposes}
\footnotetext[2]{Their mechanism can also identify unique device events}
\footnotetext[3]{A Physical Unclonable Function (PUF) is being used rather than any network traffic based approaches}
\footnotetext[4]{They also use Flow in the shape of inter-arrival time}

%% file: mtable.tex
\begin{table*}[t!]
\caption{The classification of existing IoT fingerprinting mechanisms based on seven key categories}  
\label{tab:classification}
\begin{tabular}{c|c|c|c|c|c|c|c|c}
\hline
 \makecell{Passive/\\Active\\(Cat 1)} & \makecell{Static/\\Dynamic\\(Cat 2)} & \makecell{MAC/\\Network/\\Application\\(Cat 3)} & \makecell{Packet/\\DataFrame/\\Flow\\(Cat 4)} & \makecell{Header/\\Payload\\(Cat 5)} & \makecell{Rule-based/\\ML-based\\(Cat 6)} & \makecell{Class/\\Type/\\Unique\\(Cat 7)} & Papers & \makecell{With respect to\\the first two\\main features} \\
 \hline
 \hline
 P & S & N & P & H & RB & C+T & \cite{guo2018ip} & \multirow{5}{*}{\makecell{Passive\\and\\Static}} \\
 P & S & N & P & H & ML+RB & C & \cite{desai2019feature}\footnotemark\\
 P & S & N & P+F & H & RB & T & \cite{saidi2020haystack}\\
 P & S & N & F & H & ML & C & \cite{Aneja2018}\\
 P & S & M+N & P & H+P & ML & C+T & \cite{Bezawada2018}\\
 \hline 
 P & D & N & F & - & ML & C+T & \cite{gu2018bf} & \multirow{5}{*}{\makecell{Passive\\and\\Dynamic}} \\
 P & D & N & P+F & H+P & ML & C & \cite{thangavelu2018deft} \\
 P & D & M & DF & H+P & RB & U & \cite{Siby2017} \\
 P & D & M+N & P & H+P & ML & C\footnotemark & \cite{trimananda2019} \\
 P & D & M+N+A & P+F & H+P & ML+RB & U & \cite{sivanathan2017characterizing} \\
 \hline 
 A & S & N & P & H+P & ML & C+T & \cite{yang2019towards} & \multirow{1}{*}{\makecell{Active and Static}} \\
 \hline 
 A & D & M & DF & P & ML & - & \cite{bratus2008active} & \multirow{2}{*}{\makecell{Active and Dynamic}} \\ 
 A & D & A & P & H+P & RB & C+T & \cite{Feng2018}\\ 
\hline
 P & S+D & N & P & P & ML & T & \cite{apthorpe2017closing} & \multirow{17}{*}{\makecell{Employing\\a combination of\\approaches}}\\
 P & S+D & N & P & H+P & ML & C & \cite{Nguyen2019}\\
 P & S+D & N & F & H & ML & U & \cite{meidan2017}\\
 P & S+D & N & P & H & ML & C & \cite{Bai2018}\\
 P & S+D & N & P+F & P & ML & C & \cite{Ortiz2019}\\
 P & S+D & N & F & P & ML & U & \cite{kotak2020iot}\\
 P & S+D & N & P & H+P & RB & C+T & \cite{Hafeez2020}\\
 P & S+D & N & P & H & ML & U & \cite{bremler2019iot}\\
 P & S+D & N & P+F & H & ML & C & \cite{msadek2019iot} \\
 P & S+D & N & P+F & H & ML & - & \cite{pellegrino2017learning}\\
 P & S+D & N & P+F & H+P & ML & C & \cite{marchal2019audi}\\
 P & S+D & N & P+F & H+P & ML & C & \cite{Desai2019}\\
 P & S+D & M+N+A & P+F & H+P & ML & C & \cite{Aksoy2019}\\
 P & S+D & N & P+F & H+P & ML & C& \cite{Salman2019}\\
 P & S+D & N & P+F & H+P & RB & C& \cite{Yadav2019}\\
 A & S+D & - & - & - & RB & C & \cite{marchand2017implementation}\footnotemark \\
 P+A & S+D & N & P\footnotemark & H+P & ML & C & \cite{manginointernet} \\
 P+A & D & N & F& H & ML & U & \cite{Radhakrishnan2015} \\
 \hline
\end{tabular}
\end{table*}

%% file: 5_conclusion.tex
\section{Conclusion}
\label{sec:conclusion}

IoT fingerprinting has become an important research area, due to the increased prevalence of IoT devices. Fingerprinting mechanisms serve as a key component in a network administrator's effort to identify and categorise IoT devices, in order to be able to observe and manage IoT devices in their network properly, especially in relation to pinpointing potential causes of security problems.

There are many existing IoT fingerprinting mechanisms available, but it is not easy to choose a suitable mechanism for one's network, because there is currently no consistent framework for analysing these mechanisms. This is a gap that our paper aims to address. Firstly, we compiled a list of key terminologies that are essential in understanding and analysing IoT fingerprinting mechanisms. From there, we carefully constructed a framework called \textbf{\textit{\system}}, which provides a frame of reference for a fair and consistent comparison of these mechanisms. And finally, we demonstrated the usefulness of our framework by populating a table with some example mechanisms.
We mainly focused on the mechanisms that use Machine Learning techniques. However, there are several mechanisms employing Rule-based techniques that are worth mentioning.

There are several key insights that came up from our research. We found that the majority of existing IoT fingerprinting mechanisms use a passive fingerprinting approach. This means a less intrusive approach is generally favoured. Furthermore, a dynamic approach -- or a combination of both static and dynamic approaches -- is very popular, quite possibly due to the need to fulfill a liveness property to minimise the risk of stale data or replay attacks. On the other hand, the least common mechanism seems to be based on a combination of active and static approach. This could be because such a combination might be more prone to a spoofing attack.

While we endeavoured to be as thorough and exhaustive as we can in our research, we are aware that there are some limitations in our work. For example, there are seven categories that we mainly consider in our framework, as presented in Table \ref{tab:classification}. However, it is possible that there are other categories that need to be considered in more detail. Furthermore, our current classification is mostly based on the software-related features of the IoT fingerprinting mechanisms. It would be more complete if hardware-related features -- in particular, leveraging the Physical Unclonable Function (PUF) features -- are also considered.

An interesting direction for future work is to look at the distribution and impact of IoT fingerprinting mechanisms and see if any significant patterns emerge in terms of mechanisms that are more popular or more effective, and how such patterns change over time. 

Further work will also need to be carried out to validate our framework. This could be done by utilising publicly available datasets (provided by some IoT fingerprinting mechanisms) in an experiment to classify real-world IoT devices in a live setting. This will allow for the accuracy of existing mechanisms to be calculated, enabling a fairer comparison of these mechanisms to be performed. Achieving this will enable system administrators to justify their choices with regard to which IoT fingerprinting mechanism(s) they would like to employ in their network.

\section*{Acknowledgement}
Yadav and Feraudo are supported in part by "Data Negotiability in Multi-Mode Communication Networks" sub-awarded project  funded by EPSRC grant EP/R045178/1.